\documentclass[prl,superscriptaddress,twocolumn,preprintnumbers]{revtex4}

\usepackage{color}
\usepackage[active]{srcltx}
\usepackage{amsmath,amsfonts,amssymb,amsthm,amstext,amscd,eucal,srcltx}
\usepackage{epsfig,graphicx,bm}
\usepackage{epstopdf, epsf}
\usepackage{dcolumn}
\usepackage{hyperref}

\def\bef{\begin{figure}}
\def\eef{\end{figure}}

\newcommand{\be}[1]{\begin{equation}\label{#1}}
\newcommand{\beq}{\begin{equation}}
\newcommand{\ee}{\end{equation}}
\newcommand{\beqn}[1]{\begin{eqnarray}\label{#1}}
\newcommand{\eeqn}{\end{eqnarray}}
\newcommand{\bd}{\begin{displaymath}}
\newcommand{\ed}{\end{displaymath}}

\def\lsim{\raise0.3ex\hbox{$\;<$\kern-0.75em\raise-1.1ex
e\hbox{$\sim\;$}}}
\def\gsim{\raise0.3ex\hbox{$\;>$\kern-0.75em\raise-1.1ex
\hbox{$\sim\;$}}}
\def\simlt{\mathrel{\lower2.5pt\vbox{\lineskip=0pt\baselineskip=0pt
           \hbox{$<$}\hbox{$\sim$}}}}
\def\simgt{\mathrel{\lower2.5pt\vbox{\lineskip=0pt\baselineskip=0pt
           \hbox{$>$}\hbox{$\sim$}}}}
\def\unity{{\hbox{1\kern-.8mm l}}}

\begin{document}

\preprint{IPM/P-2012/009} 
\vspace*{3mm}
\title{A new duality between Topological M-theory and Loop Quantum Gravity}

\author{Andrea Addazi}
\email{andrea.addazi@lngs.infn.it}
\affiliation{Center for Field Theory and Particle Physics \& Department of Physics, Fudan University, 200433 Shanghai, China}

\author{Antonino Marcian\`o}
\email{marciano@fudan.edu.cn}
\affiliation{Center for Field Theory and Particle Physics \& Department of Physics, Fudan University, 200433 Shanghai, China}

\begin{abstract}
\noindent
Inspired by the long wave-length limit of topological M-theory, which re-constructs the theory of $3+1$D gravity in the self-dual variables' formulation, we conjecture the existence of a duality between Hilbert spaces, the ${\bf H}$-duality, to unify topological M-theory and loop quantum gravity (LQG). By ${\bf H}$-duality non-trivial gravitational holonomies of the kinematical Hilbert space of LQG correspond to space-like M-branes. The spinfoam approach captures the non-perturbative dynamics of space-like M-branes, and can be claimed to be dual to the S-branes foam. The Hamiltonian constraint dealt with in LQG is reinterpreted as a quantum superposition of SM-brane nucleations and decays. 
\end{abstract}


\maketitle

\noindent 
M-theory and loop quantum gravity (LQG) are usually accounted as the favored candidates to solve the problem of quantum gravity. Nonetheless, both the frameworks are still concerned with several technical and conceptual issues, and the lack of any direct experimental data on the string/Planck scale only allows to focus on the mathematical self-consistency of the two frameworks. It is also commonly retained that these theories cannot be compatible with one another. Reasons to prefer M-theory to LQG are usually individuated in the unification of all particles and interactions, and in the convergence --- at the low-energy perturbative limit --- to general relativity. On the other hand, LQG has the remarkable advantage to be background independent and to provide a fully non-perturbative theory --- unification models were also proposed in LQG \cite{BilsonThompson:2006yc,Alexander:2011jf,Alexander:2012ge}. 
It seems therefore to be very challenging to reconcile theories that are so much different one another: M-theory requires supersymmetry and higher dimensions, while in LQG the phase space variables, the so called Ashtekar variables, were originally defined only in 4 dimensions and without need of supersymmetry. Nonetheless, supersymmetric and higher dimensional reformulations of LQG have been also considered in the literature, even if their dynamics was not deepened in detail as it happened for the standard formulation of LQG. Furthermore, the history of string theory, in which many different models were connected by dualities, actually suggests that string theory and LQG may be unified despite their profound differences. This urges the question: {\it can M-theory and LQG describe aspects of the same theory, but from two different point of view?}

Recently, a promising unification approach between string theory and LQG was proposed in Refs.~\cite{Turiaci:2017zwd,Mertens:2017mtv}, within the context of holographic AdS\!\!\!$\phantom{a}_{2}$/CFT and AdS\!\!\!$\phantom{a}_{3}$/CFT models. In this letter, we spell some arguments to propose a new possible correspondence principle between LQG and M-theory. This is a conjecture sustained by two insightful facts. 
First, topological M-theory on 7-dimensional $G_{2}$ manifold has a long wave-length description \cite{Dijkgraaf:2004te} in terms of 3-form $\Phi$ --- this is a result based on seminal papers by Hitchin \cite{Hitchin:2000jd,Hitchin:2001rw}. Then, thanks to a foliation of $G_{2}$ into four dimensional $M$ sub-manifold, we can demonstrate that $\Phi$ corresponds to the self-dual formulation of gravity \cite{Dijkgraaf:2004te}. This may suggest the following conjecture: the quantization procedure deployed in LQG captures aspects of the non-perturbative M-branes dynamics. In particular, the Hamiltonian formulation of LQG may provide a description for the space-like M-branes, {\it i.e.} the SM-branes. 
Second, it was noticed that the holonomy of a flat connection can be non-trivially recognized to be different than unity, if and only if a non-trivial (space-like) line defect is localized inside the loop. This is true while considering generic holonomies \cite{Bianchi:2009tj}, and it corresponds to a sort of Aharanov-Bohm effect of the self-dual gravitational field. We are tempted to suggest that this defect can be identified with space-like strings. These objects can be indeed understood as one dimensional S-branes, while these latter completely break all supersymmetric generators, eventually allowing a non-supersymmetric spinfoam approach. 

\begin{figure}[t!!]
\begin{center}
\vspace{1cm}
\includegraphics[width=7cm,height=5cm,angle=0]{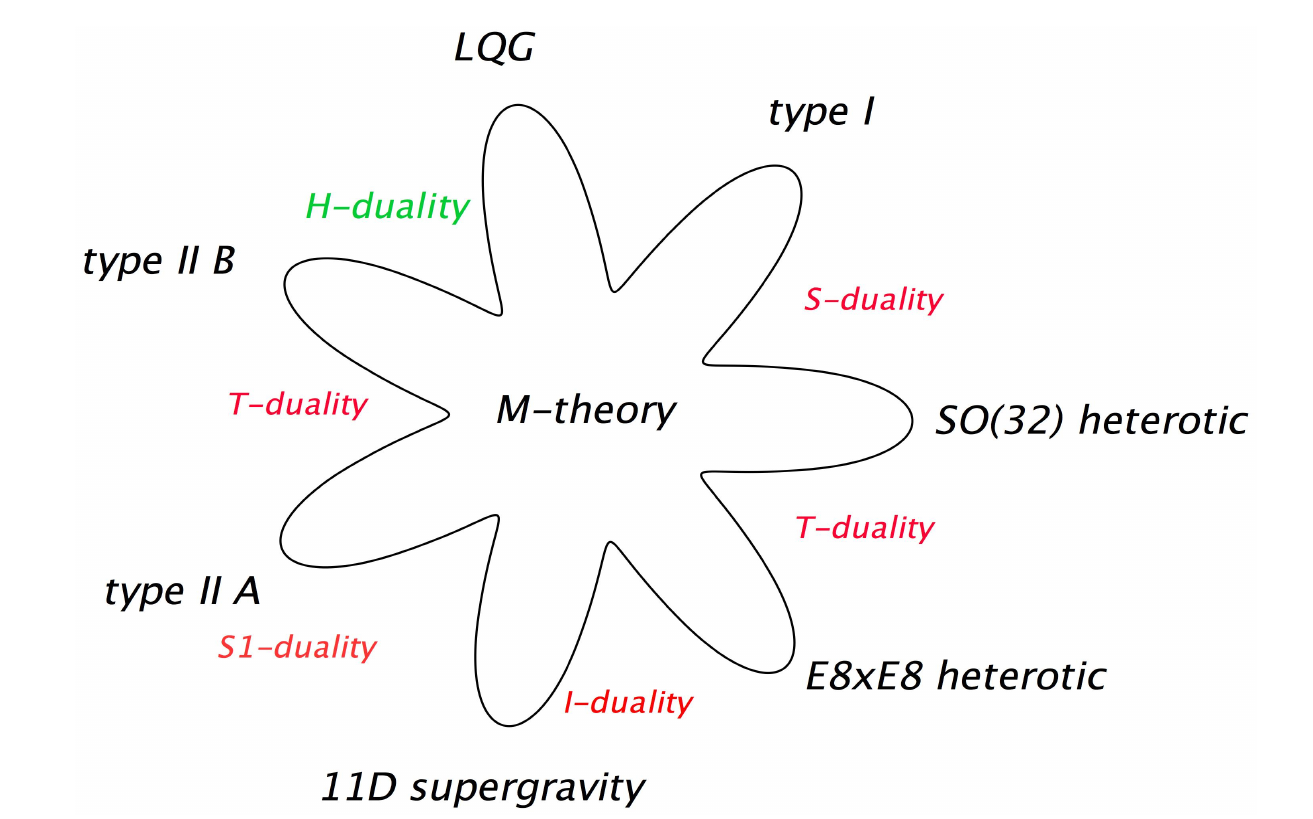}
\caption{We show the seven epiphenomena of non-perturbative M-theory. In particular, we conjecture that LQG is recovered from the duality among holonomies in the Hilbert space of LQG and space-like fundamental strings in IIB-string theory, related to SM2-branes by a double reduction. We dub this new duality ${\bf H}$-duality (written in green and located between IIB-type and LQG). Other duality transformations hitherto discovered are also represented in red. } \label{fig:1}
\end{center}
\end{figure}

For topological M-theory in a 7D manifold $\mathcal{M}$ equipped with a real three form $\Phi$, the action is described by
\cite{Hitchin:2000jd,Hitchin:2001rw}
\be{HH}
I=\int_{\mathcal{M}}\sqrt{h(\Phi)}\,,
\ee
where $\sqrt{h}h_{ab}=\Phi_{acd}\Phi_{bef}\Phi_{ghi} \epsilon^{cdefghi}$. In other words, the metric tensor can be completely recast in terms of a three form field $\Phi$. This is very much the same of what happen in LQG, where the densitized metric is cubic in the form field. Equivalently, the action can be rewritten in terms of the dual 4-form field $G$ instead of $\Phi$.

We first fix the cohomology class of $\Phi$, by setting $\Phi=\Phi^{0}+d\beta$, where $\beta$ is a two form. With such ``fixing'', which individuates a class of cohomologies, the action is invariant under gauge transformations, which are locally parameterized by a 1-form $\lambda$, {\it i.e.} $\beta \rightarrow \beta'=\beta+d\lambda$. The Hitchin action then reads 
\be{IPHI}
I[g,\Phi]=\int_{\mathcal{M}}[\sqrt{g}-g^{ab}\sqrt{h}h_{ab}]\,.
\ee
Then we can focus on $\mathcal{M}=\Sigma \times R$, where $\Sigma$ is 6D manifold. Fixing the time coordinate, we can define the canonical momenta $\pi^{ij}={\delta I}/{\delta \dot{\beta}_{ij}}$, where the dot is the usual derivative on $R$ coordinates. The primary constraints are generated by $\pi^{0i}={\delta I}/{\delta \dot{\beta}_{0i}}=0$, while the Poisson algebra is 
$$\{\beta_{ij}(x),\pi^{kl}(y)\}=\delta_{ij}^{kl}\delta^{6}(x,y).$$ 
The Hamiltonian constraints can be written as --- see {\it e.g.} Ref.~\cite{Smolin:2005gu} --- $\mathcal{H}=\mathcal{K}-c \mathcal{V}$, where $c$ is an dimensionless constant, and $\mathcal{V}=\tilde{\kappa}_{i}^{j}\tilde{\kappa}_{j}^{i}$, with $\tilde{\kappa}_{i}^{j}=\Phi_{ikl}\Phi_{mno}\epsilon^{klmnoj}$, and $\mathcal{K}=\pi^{ij} \pi^{kl} \pi^{mn} \epsilon_{ijklmn}$.

Consequently, smearing against a test function $N$, the scalar constraint becomes 
\be{HN}
\mathcal{H}(N)=\int_{\Sigma} N\, (\pi^{ij} \pi^{kl} \pi^{mn} \epsilon_{ijklmn}-a\tilde{\kappa}_{i}^{j}\tilde{\kappa}_{j}^{i})\,, 
\ee
which  closes the Poisson algebra 
\be{HNHM}
\{\mathcal{H}(N),\mathcal{H}(M)\}=\int_{\Sigma}D_{j}\omega_{NM}^{j}\, ,
\ee
where $\omega_{NM}^{j}=18a(N\partial_{i}M-M\partial_{i}N)\pi^{ik}\tilde{\kappa}_{k}^{j}$ and $a$ stands for a numerical factor, fixed to $1/4$.

There are several examples of gravity forms theories in lower than \emph{7} dimensions that can be still connected to topological M-theory, through dimensional reduction of the latter. For instance, in topological M-theory on a $G_{2}$ manifold \cite{Dijkgraaf:2004te}, a metric theory can be either reconstructed from the 3-forms $\Phi$ or dually from 4-forms $G=\star \Phi$. Thus the metric is not a fundamental field of this theory, since it can be reconstructed from the $\Phi$-field. The Einstein-Hilbert action with cosmological constant, which reads $\mathcal{S}_{\rm GR}=\int_{\mathcal{M}_3} \sqrt{-g} \, (E-2 \Lambda)$, is a topological theory, when cast in \emph{3} dimensions. As a matter of facts, the Einstein-Hilbert action rewrites in the first-order formalism as 
\begin{eqnarray} \label{3dl}
\mathcal{S}=\int_{\mathcal{M}_3} {\rm Tr}
\left( e\wedge F+\frac{\Lambda}{3} e\wedge e \wedge e \right),
\end{eqnarray}
with $F(A)=dA+A \wedge A$ the 2-form field-strength of an SU(2) gauge connection $A$
and $e$ the 1-form on $M$ valued in $SU(2)$. The metric is related to vielbiens $e_{a}^{i}$ by $g_{ab}=-{1}/{2}{\rm Tr}(e_{a}e_{b})$. The action \eqref{3dl} can reformulated as a Chern-Simons gauge theory.

The action of the 4D self-dual sector of LQG reads 
\begin{eqnarray}
S=\int_{M^{4}}\Sigma^{k}\wedge F_{k}-\frac{\Lambda}{24}\Sigma^{k}\wedge \Sigma_{k}
+\Psi_{ij}\Sigma^{i}\wedge \Sigma^{j}\,,
\end{eqnarray}
where $A^{k}$ is an $SU(2)$ gauge field with $F^{k}=dA^{k}+e^{ijk}A^{i}\wedge A^{j}$ and $\Sigma_{k}$ is a $SU(2)$ triplet of 2-forms fields $k=1,2,3$; $\Psi_{ij}$ is a scalar field on $M$ which is a symmetric representation of $SU(2)$. Varying the action with respect to $\Psi_{ij}$, we can derive 
\begin{eqnarray}
\Sigma^{(i}\wedge \Sigma^{j)}-\frac{1}{3}\delta^{ij}\Sigma_{k}\wedge \Sigma^{k}=0\,.
\end{eqnarray}
Such a condition implies that $\Sigma^{k}$ can
be re-expressed in terms of the vierbein $\Sigma^{k}=-\eta_{ab}^{k}e^{a}\wedge e^{b}$, where $e^{a}$ is the vierbien 1-forms on $M^{4}$, $a=1,...,4$ and $\eta_{ab}^{k}$ is the 't Hooft symbol $\eta_{ab}^{k}=\epsilon_{ab0}^{k}+\frac{1}{2}\epsilon^{ijk}\epsilon_{ijab}\,.$ As renown, the vierbein is in turn related to the metric by the relation $g=\sum_{a=1}^{4}e^{a}\otimes e^{a}$. The two-forms $\Sigma^{k=1,2,3}$ are self-dual with respect to the metric $g$, {\it i.e.} $\Sigma^{k}=*\Sigma^{k}$. One can also rewrite the metric directly in terms of $\Sigma$, finding 
\begin{eqnarray}
\sqrt{g}g_{ab}=-\frac{1}{12}\Sigma_{aa_{1}}^{i}\Sigma_{ba_{2}}^{j}\Sigma_{a_{3}a_{4}}^{k}\epsilon^{ijk}\epsilon^{a_{1}a_{2}a_{3}a_{4}}\,.
\end{eqnarray}

Local models of a complete 7-manifold $X$ can be obtained as a m-dimensional vector bundle on an n-dimensional cycle $M$, such that $m+n=7$. Local gravitational modes induce a lower-dimensional gravity on $M$. The equations of motion of topological M-theory 
\begin{eqnarray}
d\Phi=0\,, \qquad d_{\star\Phi}\Phi=0\, ,
\end{eqnarray}
lead to the equations of motion of the p-form fields on $M$, in turn interpreted as topological gravity equation of motion  on $M$. $\Phi$ can be decomposed as a combination of vielbein components $e^i$. The equation $d\Phi=0$ is equivalent to the equations of motion of 3D gravity for 3D fiber, and yields $de=-A\wedge e-e\wedge A$ and $dA=-A\wedge A-\Lambda e\wedge e$, which are precisely equivalent to the equations of motion of 3D Chern-Simons gravity.
The latter also fulfills $d_{\star\Phi}\Phi=0$. The $\Phi$-field can be rewritten as a combination of vielbein as
\be{PhiA}
\Phi=Ae^{123}+Be_{i}\wedge \Sigma^{i}\,,
\ee
where $\Sigma^{1}=\alpha^{12}-\alpha^{34}$ and so on for the other components, $\alpha_{i}$ denoting a 1-form in the fiber direction, {\it i.e.} $\alpha^{i}=D_{A}y^{i}=dy^{i}+(Ay)^{i}$.

Coming back to the reduction to the 4D gravity, we can decompose $\Phi$ as
\be{Phialphap}
\Phi=\alpha^{123}+\alpha^{1}\wedge \Sigma^{1}
+\alpha^{2}\wedge \Sigma^{2}+\alpha^{3}\wedge \Sigma^{3}\,.
\ee
In Ref.~\cite{Smolin:2005gu}, Smolin suggested a quantization scheme for these theories, defining the holonomy of the 1-form $\beta$ and its conjugate variables respectively as $H[S]=e^{\int_{S}\beta}$ and the momentum flux operators $\Pi[A]=\int_{A}\pi^{*}$.
The Poisson brackets can be then recovered as
\be{TSA}
\{H[S],\Pi[A]\}=I[S,A]H[S]\, ,
\ee
where $I[S,A]$ is the intersection number of the surfaces $S,A$. This allows to define networks $\Gamma$ on the two surfaces, with their relative Hilbert states such that $\langle \Gamma|\Psi\rangle=\Psi(\Gamma)$.\\
 
The second insightful fact we have been inspired by in stating the correspondence principle between LQG and M-theory, is the reformulation, suggested in Ref.~\cite{Bianchi:2009tj}, of states and scalar products of LQG in terms of non-trivial holonomies enclosing defects. One can start from a 3-manifold $\Sigma$ with a network of defect-lines. A locally-flat connection on the spatial 3-manifold can have non-trivial holonomy, as in the electromagnetic Aharanov-Bohm effect. Quantizing the theory, Bianchi obtained in Ref.\cite{Bianchi:2009tj} a scalar product that is the same used in LQG. We then consider a flat connection in $\Sigma'=\Sigma-l$, where $l$ is a defect line, and then the holonomy encircling this line. The induced metric on $\Sigma$ is $q_{ab}(x)$, which allows to choose the Coulomb-gauge as $\chi=q^{ab}\partial_{a} A_{b}$. The line is fixed along the z-axes in Euclidean metric. Moving from the gauge fixing condition $\chi^{i}\!\!=\!\!\partial^{a}A_{a}^{i}\!\!=\!\!0$, the Ashtekar connection reads $A_{a}^{i}={f^{i}}/{2\pi}\alpha_{a}(x)$, in which $\alpha_{a}(x)\!\!=\!\!\left(-{y}/({x^{2}+y^{2})},{x}/({x^{2}+y^{2})},0\right)$. The associated holonomy along the loop $\gamma$ is then $h_{\gamma}[A]={\rm exp}\left[i\left(\int_{\gamma}\alpha_{a}dx^{a}\right){f^{i}}/{2\pi}\tau_{i}\right]$, while the related non-abelian magnetic field recasts $B_{i}^{a}\equiv \frac{1}{2}\epsilon^{abc}F_{bc}^{i}=\int_{l}ds f_{i} \dot{x}^{a}(s)\delta^{(3)}(x-x(s))$. The flux of the magnetic field through the surface $S$ punctured by the curve $\gamma$ reads 
\be{MFF}
\mathcal{F}_{i}[B,S]=\int_{S} B_{i}^{a}(X(\sigma))\epsilon_{abc}\frac{\partial X^{b}}{\partial \sigma^{1}}\frac{\partial X^{c}}{\partial \sigma^{2}}d\sigma^{1}d\sigma^{2}
\ee
$$=\int_{l} ds \int_{S}d\sigma^{1}d\sigma^{2}f_{i}\epsilon_{abc}\dot{x}^{a}(s)\frac{\partial X^{b}}{\partial \sigma^{1}}\frac{\partial X^{c}}{\partial \sigma^{2}}\delta^{(3)}(X(\sigma)-x(s))\,, $$
which is just equal to $f_i$, the flux of the magnetic field through the defect line. This is analogous to the problem of a cylindrical solenoid in electromagnetism. The related moduli space is $\{ f^{i}\in S_{3}\}/SU(2)=\{\phi \in [0,2\pi]\}$, where $\Psi^i$ depends only on the moduli $\phi$, while is invariant under global $SU(2)$ rotations.

In this framework, the scalar product of states depending by the moduli space can be put in correspondence with the LQG scalar product of holonomy states in the Hilbert space $\mathcal{K}'$. We then find $$\langle g|g' \rangle=\int_{\mathcal{A}_{f}/\mathcal{G}}\mathcal{D}[A]\bar{\Psi}_{g}[A]\Psi_{g'}[A]=\int_{\mathcal{N}}\prod_{r}dm_{r}J\Delta_{FP}\bar{g}g'\,,$$
where $J$ denotes the Jacobian and $\Delta_{FP}$ the Faddeev-Popov determinant. In order to prove such equivalence, one may compute it directly in the case of one line defect. 
The Jacobian in spherical coordinates reads  $J(\phi)=\phi^2$, which is associated to $d^{2}\Phi=\phi^{2}d\phi d^{2}v^{i}$. The Faddeev-Popov term $\Delta_{FP}$ is given by the determinant of the operator $K$, which is the derivative of the gauge fixing condition $\chi^i\!\!=\!\!0$ with respect to the gauge parameter, namely $K=-\delta_{ij}\Delta-\epsilon_{ijk}\frac{\Phi^{k}}{2\pi}\alpha^{a}\partial_{a}$. Its eigenvalues are $\lambda_{n}=n^{2}+n{\phi}/{2\pi}$, where $n=\pm 1,\pm 2,\dots$ (twice degenerate). Using the expression for the (regularized) Faddeev-Popov determinant \cite{Bianchi:2009tj}, we obtain $\langle g|g'\rangle=\frac{1}{\pi}\int_{0}^{2\pi} \sin^{2}(\phi/2) \bar{g}g'$. This can be compared to the scalar product of LQG. A natural choice in LQG is the Haar measure on the links of the graphs, which reads $\langle \eta|\zeta\rangle=\int_{\mathcal{A}_{l}/\mathcal{G}}\mathcal{D}[A]\bar{\Psi}_{\Gamma,\eta}[A]\bar{\Psi}_{\Gamma',\zeta}[A]$. After few steps we can reshuffle the Haar measure as $\langle \eta|\zeta\rangle\!\!=\!\!\int_{SU(2)^{L}}\prod_{l=1}^{L}d\mu_{H}(h_{l})\bar{\eta}(h_{1},...,h_{L})\zeta(h_{1},...,h_{L}),$ in terms of the class of graphs $\Gamma'$ dual to the cellular decomposition. Using the Peter-Weyl theorem with this choice of the scalar product, the spin-network states, with graph $\Gamma'$, provide an orthonormal basis of the Hilbert space $\mathcal{K}'$. The spin-network basis entails cylindrical functions $\eta(h_{1},...,h_{L})$, the holonomies of which are labeled by $SU(2)$ representations. To every node of the spin-network states are assigned intertwiners that realize an invariant map onto the tensor product of the representations. In particular, we can recognize that 
$\eta_{j_{1}j_{n}}(h_{1},...,h_{L})\!\!=\!\!\left(\bigotimes_{n}v_{in}\right)\!\!\left(\bigotimes_{\gamma_{l}}D^{(j_{l})}(h_{L})\right)$,
where $n \,\!\!\in \,\!\!\Gamma'$, $\gamma_{l}\!\!\in\!\! \Gamma'$, in such a way that the orthonormality of the spin-network states reads $\langle \eta_{j_{1}i_{n}}|\eta_{j'_{l}i'_{n}}\rangle\!\!=\!\!(\prod_{l} \delta_{j_{l}j_{l}'})(\prod_{n}\delta_{i_{n}i_{{n}'}})$. The states $\Psi_{\gamma,\eta}[A]=\eta(h_{\gamma}[A])$ are then individuated by a complex-valued function $\eta$ on $SU(2)$, and by the homotopy class $[\gamma]$ of loops closing one time the defect $l$. The scalar product can be cast in terms of the Haar measure on $SU(2)$. In particular, if we define $f(\phi)=\eta(e^{i\phi\tau_{3}})$, we obtain $\langle \eta|\zeta \rangle=\langle g|g'\rangle$.\\

\begin{figure}[t!!]
\begin{center}
\vspace{1cm}
\includegraphics[width=6cm,height=6cm,angle=0]{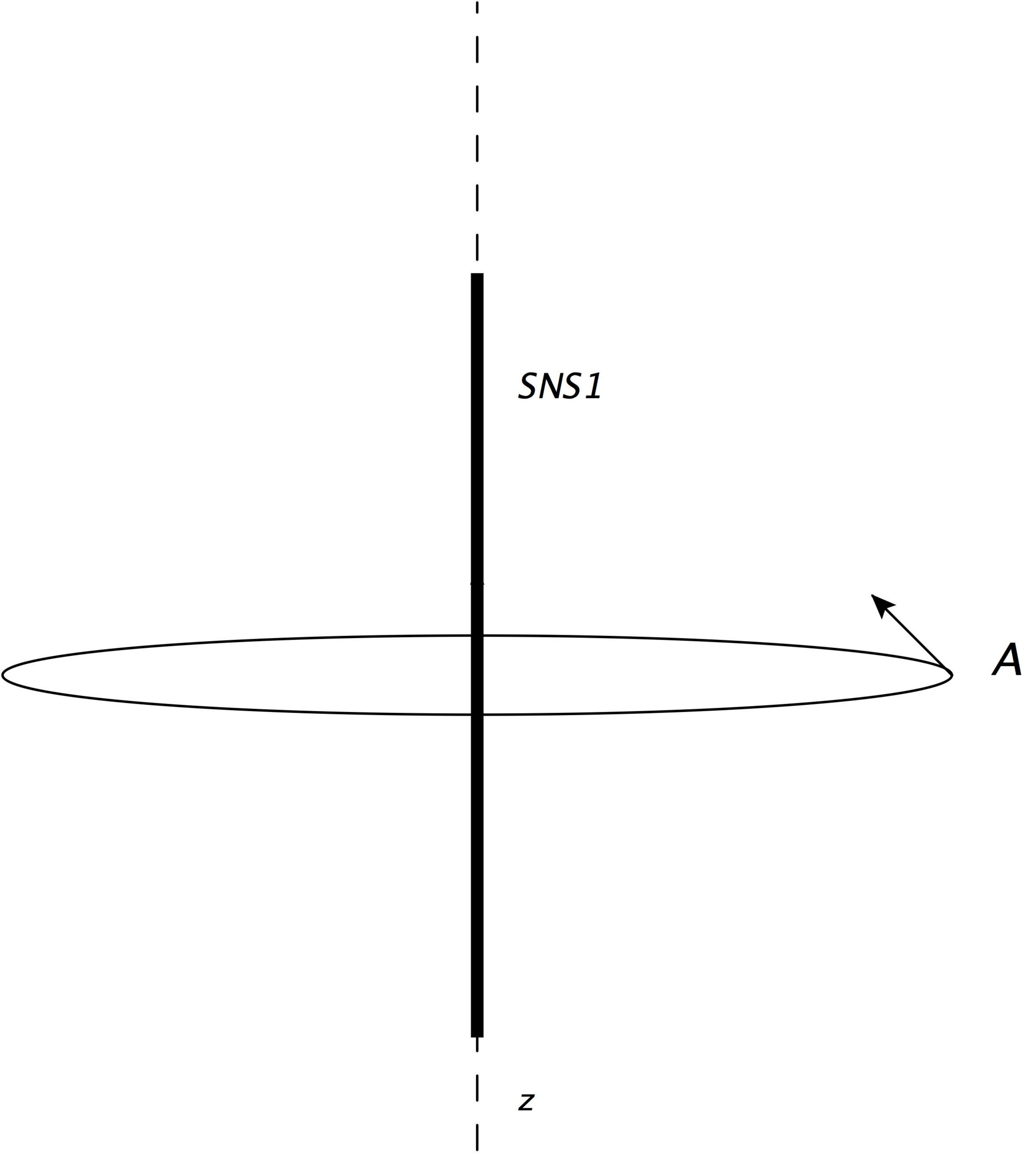}
\caption{The presence of the space-like fundamental string is associated to a non-trivial holonomy of self-dual variables. } \label{fig:1}
\end{center}
\end{figure}

We can generalize the example provided by one line defect, to the case of a network $\mathcal{S}$ of curves in $\Sigma$ \cite{Bianchi:2009tj}. Let us define a locally-flat connection $A(x)$. The holonomies of these connections are not trivial, since they encircle the net of defects. Let us consider, the space of locally flat connections, modulo the gauge transformations, which we call $\mathcal{A}_{f}/\mathcal{G}$. Now, the configuration in such a space  corresponds to a homomorphism from $\pi_{1}(\Sigma-\mathcal{S})$ into $G$, cosetting gauge transformations. This defines the moduli space $\{m_{r}\}$, contained in $\{m_{r}\}\equiv\mathcal{N}\equiv {\rm Hom}\{\pi_{1}(\Sigma-\mathcal{S},G)/G\}$. In this generalized set-up, the states can be put in correspondence with functions of the moduli $\Psi_{\Gamma,\eta}[A^{m_{r},g}]=f(m_{1},...,m_{r})$, where $\Psi_{\Gamma,\eta}[A]=\eta(h_{\gamma_{1}}[A],...,h_{\gamma_{L}}(A))$ and $\eta$ is a complex function valued in $SU(2)$, {\it i.e.} $\eta:SU(2)^{L}\rightarrow C$. The scalar products of LQG and the moduli functions are in correspondence by means of $\langle f|g\rangle =\int_{\mathcal{A}_{f}/\mathcal{G}}\mathcal{D}[A]\bar{\Psi}_{f}[A]\Psi_{g}[A]$, which reshuffles as $\langle f|g\rangle=\int_{\mathcal{N}}d\mu(m_{r})\bar{f}(m_{1},...,m_{R})g(m_{1},...,m_{R}).$

The line defect can be interpreted as a soliton charged in the self-dual gravitational gauge group --- see Fig.2. Since the scalar constraint was not yet solved in full generality, we consider space-like solitons not propagating in the time direction. They can be either space-like strings or holonomies around a circular solenoid that take a circular path orthogonal to it. 

Our conjecture is that the line defects correspond to SM-branes compactified into space-like fundamental strings, {\it i.e.} space-like NS1-branes. Fundamental S-strings are serious candidates to be considered in order to instantiate this correspondence. These are charged indeed with respect to the $\Phi$-field and consequently with respect to self-dual gravitational potential $A_{\mu}^{i}$, which contains the self-dual gravitational algebra structure. In particular, taking an holonomy $h_{\gamma}[A_{\mu}^{i}]$ encircling the fundamental S-strings, the magnetic flux on a surface $S$ punctured by the curve $\gamma$ is non-zero because essentially it surrounds a localized magnetic field form $G=\star \Phi$. There is a possible issue for this correspondence's framework: different S-branes networks can correspond to the same spin-networks and {\it viceversa}. In other words, the correspondence can be established from classes of S-branes networks to classes of LQG states. Nonetheless, the classes' correspondence is enough to guarantee that every possible S-branes system has a proper state in the kinematical Hilbert space of LQG. 

In conclusion, we conjectured the existence of a ${\bf H}$-duality, which may unify topological M-theory and LQG, arguing that non-trivial gravitational holonomies can be put in correspondence with space-like M-branes. We grounded our proposal on the long wave-length limit of topological M-theory, showing how this latter re-constructs the theory of 3+1D gravity in the self-dual variables formulation. In our considerations, we have mainly discussed the ${\bf H}$-duality between space-like NS1 foam and spinfoam, focusing more on the consequences of canonical quantization's techniques. Nonetheless it is still rather unclear whether our arguments can be more generically extended to ordinary M-branes, D-branes and NS-branes. In principle, it sounds reasonable to extend the duality and account for covariant quantization techniques. A crucial problem is to understand how S-branes and D-branes fit in this picture. From the perturbative string theory point of view, S-branes and D-brane undergo long-range interactions, exchanging closed strings. In the non-perturbative regime, this should correspond to an exchange of dilatons, gravitons and B-forms's excitations, entailing an infinite number of loops corrections. Furthermore, we should take into account also processes of brane instabilities, back-reacting on the system. We conjecture that these interactions are already encoded in the full non-perturbative regime realized on the LQG side.\\

{\bf Acknowledgments.}\\ 
We are indebted to J.~Lewandowski and H.~Verlinde for enlightening discussions during Loops'17 in Warsaw, Poland.

\end{document}